\documentclass[prx,showpacs,twocolumn,superscriptaddress,aps]{revtex4-1}

\usepackage[utf8]{inputenc}
\usepackage{graphics}
\usepackage{color}
\usepackage{stfloats}
\usepackage{amssymb}
\usepackage{amsmath}
\usepackage{overpic}
\usepackage{booktabs, array, mathptmx, float, tabularx, booktabs, lipsum, amsmath,multirow}
\usepackage{dcolumn}
\usepackage{epstopdf}
\usepackage{booktabs}
\usepackage{bm}
\usepackage{graphicx}
\usepackage{subfigure}
\usepackage{threeparttable}
\usepackage{multirow}

\begin{document}

\title{JAMIP: an artificial-intelligence aided data-driven 
infrastructure for computational materials informatics}

\author{Xin-Gang Zhao}
\altaffiliation{These authors contributed equally}
\affiliation{State Key Laboratory of Integrated Optoelectronics, College of Materials Science and Engineering, Jilin University, Changchun 130012, China}

\author{Kun Zhou}
\altaffiliation{These authors contributed equally}
\affiliation{State Key Laboratory of Integrated Optoelectronics, College of Materials Science and Engineering, Jilin University, Changchun 130012, China}

\author{Bangyu Xing}
\altaffiliation{These authors contributed equally}
\affiliation{State Key Laboratory of Integrated Optoelectronics, College of Materials Science and Engineering, Jilin University, Changchun 130012, China}

\author{Ruoting Zhao}
\altaffiliation{These authors contributed equally}
\affiliation{State Key Laboratory of Integrated Optoelectronics, College of Materials Science and Engineering, Jilin University, Changchun 130012, China}

\author{Shulin Luo}
\affiliation{State Key Laboratory of Integrated Optoelectronics, College of Materials Science and Engineering, Jilin University, Changchun 130012, China}

\author{Tianshu Li}
\affiliation{State Key Laboratory of Integrated Optoelectronics, College of Materials Science and Engineering, Jilin University, Changchun 130012, China}

\author{Yuanhui Sun}
\affiliation{State Key Laboratory of Integrated Optoelectronics, College of Materials Science and Engineering, Jilin University, Changchun 130012, China}

\author{Guangren Na}
\affiliation{State Key Laboratory of Integrated Optoelectronics, College of Materials Science and Engineering, Jilin University, Changchun 130012, China}

\author{Jiahao Xie}
\affiliation{State Key Laboratory of Integrated Optoelectronics, College of Materials Science and Engineering, Jilin University, Changchun 130012, China}

\author{Xiaoyu Yang}
\affiliation{State Key Laboratory of Integrated Optoelectronics, College of Materials Science and Engineering, Jilin University, Changchun 130012, China}

\author{Xinjiang Wang}
\affiliation{State Key Laboratory of Integrated Optoelectronics, College of Materials Science and Engineering, Jilin University, Changchun 130012, China}

\author{Xiaoyu Wang}
\affiliation{State Key Laboratory of Integrated Optoelectronics, College of Materials Science and Engineering, Jilin University, Changchun 130012, China}

\author{Xin He}
\affiliation{State Key Laboratory of Integrated Optoelectronics, College of Materials Science and Engineering, Jilin University, Changchun 130012, China}

\author{Jian Lv}
\affiliation{State Key Laboratory of Superhard Materials, College of Physics, Jilin University, Changchun 130012, China}

\author{Yuhao Fu}
\email{fuyuhaoy@gmail.com}

\affiliation{State Key Laboratory of Superhard Materials, College of Physics, Jilin University, Changchun 130012, China}

\author{Lijun Zhang}
\email{lijun_zhang@jlu.edu.cn}
\affiliation{State Key Laboratory of Integrated Optoelectronics, College of Materials Science and Engineering, Jilin University, Changchun 130012, China}

\date{\today}

\begin{abstract}
Materials informatics has emerged as a promisingly new paradigm for accelerating materials discovery and design. 
It exploits the intelligent power of machine learning methods in massive materials data from experiments or simulations to seek new materials, functionality, and principles, etc. 
Developing specialized facilities to generate, collect, manage, learn, and mine large-scale materials data is crucial to materials informatics. 
We herein developed an artificial-intelligence-aided data-driven infrastructure named Jilin Artificial-intelligence aided Materials-design Integrated Package (JAMIP), which is an open-source Python framework to meet the research requirements of computational materials informatics. 
It is integrated by materials production factory, high-throughput first-principles calculations engine, automatic tasks submission and monitoring progress, data extraction, management and storage system, and artificial intelligence machine learning based data mining functions. 
We have integrated specific features such as an inorganic crystal structure prototype database to facilitate high-throughput calculations and essential modules associated with machine learning studies of functional materials. 
We demonstrated how our developed code is useful in exploring materials informatics of optoelectronic semiconductors by taking halide perovskites as typical case. 
By obeying the principles of automation, extensibility, reliability, and intelligence, the JAMIP code is a promisingly powerful tool contributing to the fast-growing field of computational materials informatics. 
\end{abstract}

\maketitle

\section{Introduction} 
Currently, materials science research is entering a new paradigm featuring materials informatics, which applies machine-learning techniques to massive materials data\cite{agrawal_perspective_2016, schleder_dft_2019, draxl_big_2019, himanen_data-driven_2019} to accelerate materials discovery and design. 
It was first promoted in part by the Materials Genome Initiative\cite{kalil2011materials} and emerged in practice as the result of the fast development of experimental materials synthesis, characterization approaches, and theoretical materials simulation through available computation resources that yield massive materials data\cite{agrawal_perspective_2016}. 
By utilizing the recognized predictive power of materials informatics, encouraging breakthroughs have been made, including the design of new materials, the identification of new relationships\cite{ghiringhelli_big_2015, duvenaud_convolutional_2015}, and the prediction of new principles\cite{schmidt_distilling_2009, rudy_data-driven_2017}. 
In the fields of both experimental and computational materials informatics, there is an urgent need to develop powerful facilities/infrastructures to meet research requirements under this new paradigm.

Generating massive materials data of the composition-structure-property relationship is essentially the first step for materials informatics studies. 
In the field of computational materials science, there are generally three materials data generation techniques: The first is through direct calculations of single or a few materials. 
This is used frequently in the old paradigm when theorists want to study some material properties observed experimentally or to explore new physics in some previously unstudied materials. Such data are dispersedly distributed in literature and may be gathered manually or by automatic text extraction approaches\cite{fundel_relexrelation_2007, tshitoyan_unsupervised_2019}. 
The second is through the high-throughput (HT) calculations where extremely large numbers of materials spanning different chemical compositions are automatically calculated based on prototype structures. 
Along this direction, a series of computational infrastructures are developed and used in high-throughput calculations\cite{curtarolo_aflowliborg_2012, zhao_design_2017,luo_high-throughput_2021, curtarolo_aflow_2012, pizzi_aiida_2016, mathew_mpinterfaces_2016, jain_fireworks_2015, yang_matcloud_2018, zhu_sehc_2020, mathew_atomate_2017, choudhary_joint_2020, wang_alkemie_2021}. 
The accumulated computational materials data lead to the formation of the databases that are accessible for query and research purposes\cite{pickard_ab_2011, hachmann_harvard_2011, kirklin_open_2015, draxl_nomad_2019, 2016gorai}. 
The third is through crystal structure searches. These seek new stable and metastable materials structures under the typical chemical composition. 
This method complements high-throughput calculations. 
It explores the potential energy profile of materials, usually by combining artificial intelligence global optimization methods with energetic calculations\cite{glass_uspexevolutionary_2006, lonie_xtalopt_2011, pickard_ab_2011, wang_calypso_2012, tipton_grand_2013}. 
Because in structure search studies, the object of greatest concern is the lowest-energy ground-state structure, the generated materials data are usually regarded as intermediate products, and do not accumulate in materials informatics studies.

Machine learning techniques act as an engine that triggers/accelerates materials discovery in computational materials informatics\cite{schmidt_recent_2019, liu_materials_2017, ramprasad_machine_2017}. 
Various machine-learning models are already available for use in computational material informatics, such as supervised learning, unsupervised learning, and active learning\cite{kusne_--fly_2020}. 
Supervised learning requires an artificially constructed training process, which then understands the mapping relationships between material descriptors and material properties. 
Using this model, both the underlying physical laws between descriptors and the material properties can be mined\cite{ouyang_sisso_2018, bartel_new_2019}. 
Additionally, it can be used to reverse-engineer new materials to uncover new materials\cite{lu_accelerated_2018}. 
Unsupervised learning, by contrast, is generally used to uncover differences among materials or chemical systems in terms of unmarked ``descriptors''. 
Clustering algorithms are the most widely used for this purpose. 
For example, Chen et al. used the K-mean clustering algorithm to analyze the oxygen diffusion pattern, hopping statistics, and site occupation within crystals\cite{chen_molecular_2015}. 
In addition, in the field of materials discovery, active learning and Bayesian optimization, with the self-optimizing characteristics of model, allow us to search the potential materials space for high-quality materials in a limited data range, and contribute to the material experimental and computational design\cite{tran_active_2018}.

The software/infrastructure for computational materials informatics needs to meet at least the following requirement, but is not limited to them: (1) Adaptability for diverse materials discovery and design: The software should flexibly deal with different material systems from inorganic and organic to organic-inorganic hybrid systems, and different materials variations/combinations, such as alloy or defect, surface or interface, and heterostructure or superlattice. 
To fit the potential complexity of different materials with different properties, computational workflow modules should have high scalability and ensure flexible combinations among different types of computing tasks. 
(2) Efficiency and reliability in data generation and management: A high-level automation framework to initialize, run, and analyze large-scale high-throughput calculations is essential. 
Monitoring computational tasks in real-time and reporting/correcting potential errors are important. 
The software should have tools to efficiently extract and collect results and store them in a self-contained database. 
(3) Orientation by materials functionality: The software should be oriented by the functionality of materials (\textit{e.g.}, photovoltaic, thermoelectric, ferroelectric, and catalytic), which is central to materials discovery. 
Rational design of workflow modules needs to be sufficiently considered to effectively calculate the functionalities of different material systems. 
(4) Synthetical integration of data generation, storage, processing, and learning: It is desirable to integrate data generation, storage, processing, and learning modules with a unified infrastructure framework, which greatly facilitates the data fluxion and conversion procedure, as well as the efficiency of data mining. 
In the integrated framework, newly calculated data motivated by feedback from data-learning procedures may accelerate knowledge accumulation.

In this paper, we report on an artificial-intelligence-aided data-driven infrastructure that we have been developing to fulfill the above requirements for computational materials informatics. 
It is an open-source Python-integrated framework named the Jilin Artificial-intelligence-aided Materials-design Integrated Package (JAMIP). 
The code is integrated by intimately connected units of Data generation (\textit{e.g.}, high-throughput materials calculations and screening), Data collection (\textit{e.g.}, automatic data extraction, management, and storage) and Data learning (\textit{e.g.}, integrated descriptors engineering, and machine learning based data mining functions). 
Below, we describe the JAMIP code framework and its usage in carrying out materials-informatics-related research on optoelectronic semiconductors by taking halide perovskites as instances. 
The code and manuals describing its detailed usage options are freely available for download for academic research.

\section{Code Framework}
\begin{figure}[h]
\centering
\includegraphics [width=\columnwidth] {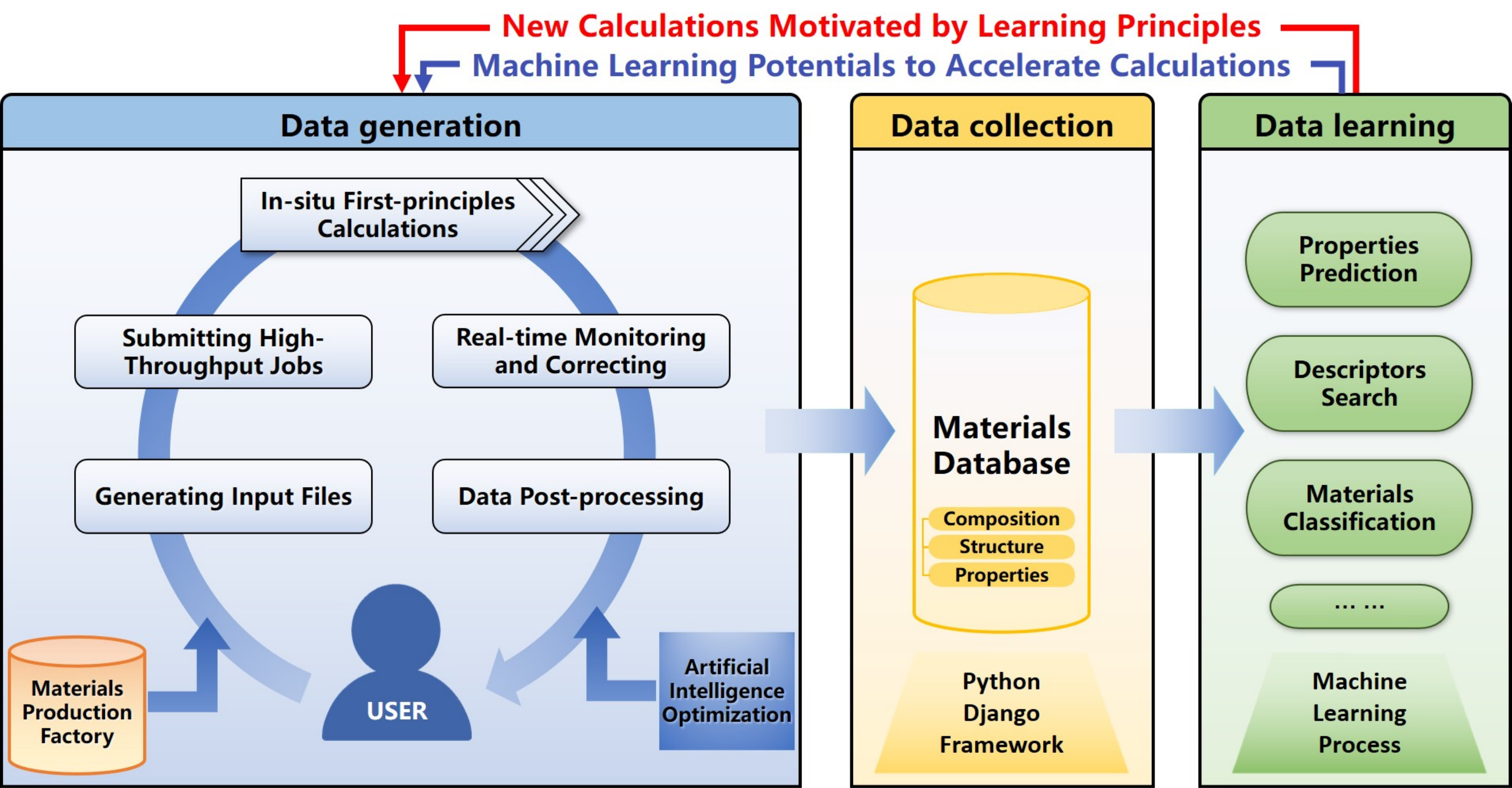}
\caption{Overview of the JAMIP code framework. 
The program comprises three major parts based on the material data's lifecycle: Data generation(blue), Data collection(yellow), and Data learning(green).}
\label{Fig1}
\end{figure}

The organization of JAMIP abides by the data lifecycle in the research field of computational materials informatics, from data generation to collection and learning, as shown in Fig.\ \ref {Fig1}. 
In the data generation stage, users perform the large-scale high-throughput materials calculations, which are done through obtaining structure prototypes from the JAMIP database, generating candidate structures through the materials production factory, initializing the input parameters of HT calculation tasks, submitting and monitoring tasks, and post-processing results. 
The code contains a specially designed computational tasks allocation and monitoring system to guarantee efficient and reliable large-scale calculations. 
We have designed packaged material property calculation modules ranging from geometrical, thermodynamic, mechanical, electronic, phononic, photonic to magnetic properties. 
There are modules for material property optimization calculations guided by artificial intelligence optimization methods (\textit{e.g.}, genetic algorithm and particle swarm optimization algorithm) to achieve optimal functionality of materials with minimal effort and global sampling in the materials space. 
Users can conveniently perform functionality-direct materials screening at this stage. 
In the data collection stage, the code has continuously developed automatic toolkits for effective extraction, collection, and analysis on target data of functional materials. 
There is a self-contained professional database for automatically normalizing and storing the ``composition-structure-property'' data of materials. 
The database is developed based on the Django framework that facilitates access to data without SQL query language. 
In the final data learning stage, the uses can exploit machine-learning methods to perform data mining researches. 
Our data learning module implements the popular general machine learning workflows, such as properties prediction, descriptors search, and materials classification. 
The learned principle knowledge from the data learning stage can in turn motivate and guide new calculations in the first data generation stage. 
The developed machine learning potentials can meanwhile accelerate high-throughput materials calculations.
 
2.1 Data generation

\begin{figure}[hb]
\centering
\includegraphics [width=\columnwidth] {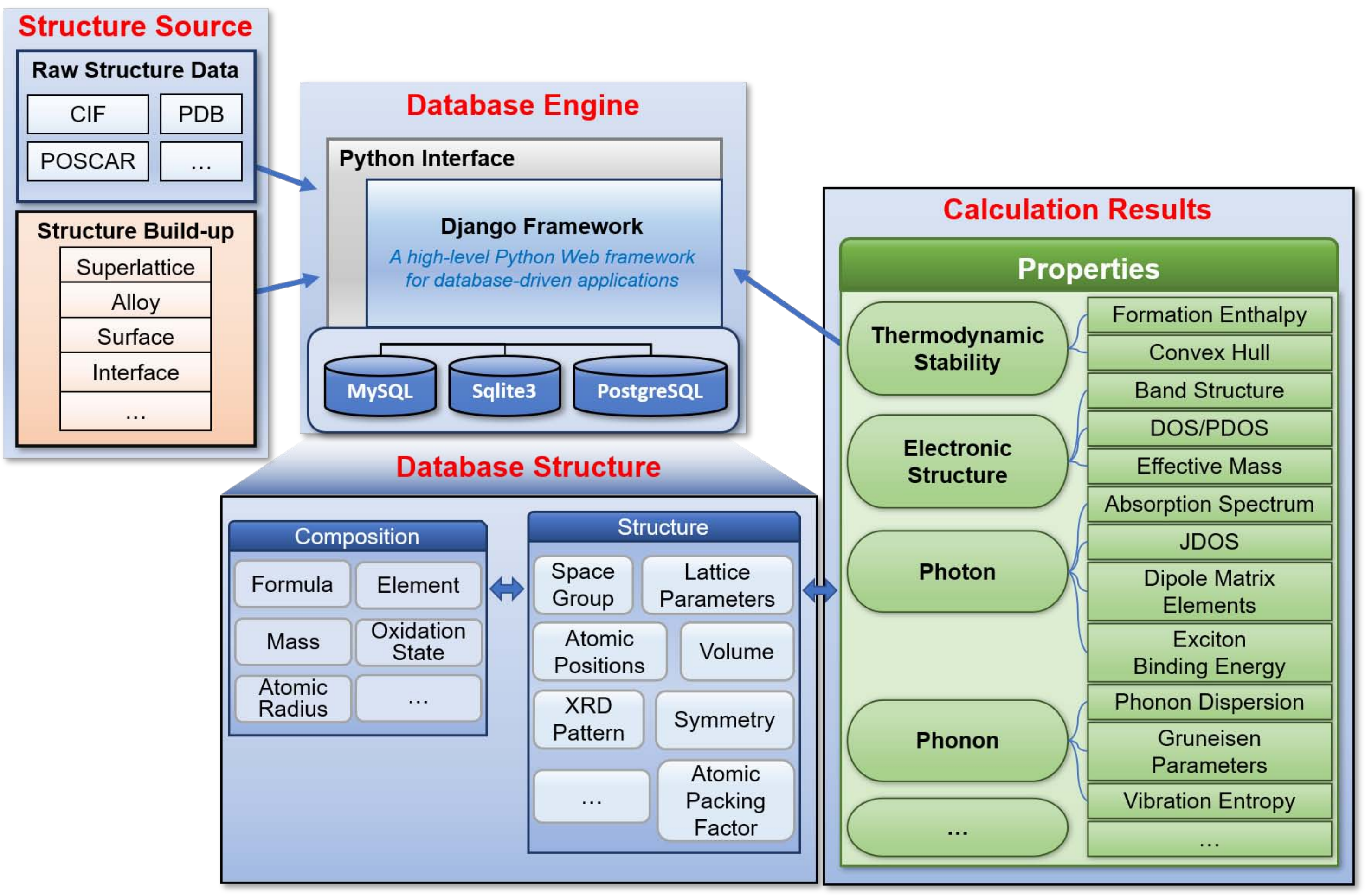}
\caption{Detailed workflow of the HT calculations.}
\label{Fig2}
\end{figure}

Materials data generation involves the configuration and initialization of the computation environment and the customization, execution, and monitoring of the calculation tasks (Fig. \ref{Fig2}).

2.1.1 Materials production factory

Currently, complex material structure manipulation mainly relies on third-party graphical interactive software. 
Although such software facilitates the users'  intuitive operation of the material structure, it also adds some obstacles when performing structural transformation in large quantities. 
Another tricky problem is the lack of available interfaces between graphical interactive software and HT software. 
With the increasing complexity of the scientific issues studied, the demand for complex structural modeling capabilities is increasing for HT calculations. 
Thus, powerful structural modeling capabilities are an important feature of next-generation HT software. 
These include building complex twin structures and structures with defects (\textit{e.g.}, plane or screw dislocations), and even generating new structures by using some structural units or symmetry operation methods based on group theory.

JAMIP has two kinds of material structure manipulation operators. 
One is a set of basic material structure manipulation operators (\textit{e.g.}, deleting atoms, adding atoms, scaling lattice parameters, supercells, defining crystal structure, and adding vacuum layer); another is a set of high-level complex material structure manipulation operators. 
They are from a flexible combination of the above basic manipulation operators (\textit{e.g.}, substituting an atom with a given molecule and constructing a crystal structure with twin interfaces). 
Users can easily further expand and customize the material structure manipulation operators based on the built-in manipulation operators.

The other important feature of our materials production factory is the integration of our developed inorganic crystal structure prototype database (ICSPD) as a sub-database of JAMIP's database. 
The structure prototype database can be beneficial for high-throughput calculations where usually tremendous materials spanning different chemical compositions are automatically calculated based on the prototype structures\cite{luo_high-throughput_2021}. 
Our ICSPD was developed in terms of the local atomic environments (LAEs) of all experimentally known inorganic crystal structures, with the help of the unsupervised learning algorithm, \textit{i.e.}, hierarchical clustering. 
It is distinct from the common structure prototypes database that was constructed directly based on the space group information of crystal structures. 
It provides a more reasonable and flexible classification based on microscopic atomic crystal structure features. 
This self-contained structure prototype database effectively facilitates ongoing high-throughput calculations in JAMIP.

2.1.2 High-throughput calculation workflow

HT calculations involve the transfer of a large number of different types of data. 
This requires the workflow module to correctly transfer data after combining different calculation tasks. JAMIP includes a task pool to manage all calculations tasks in a unified manner and transfer the data with a function called ``calculation''. 
By using the observer patterns, the task can monitor the status of tasks on the computing nodes and submit new calculation tasks in time to avoid lags in information updates and long periods of idleness. 
Furthermore, some types of calculation tasks, such as phonons, can generate multiple parallel calculations concurrently. 
JAMIP can even split these calculations into multiple compute nodes to speed up the calculations.

Scalability is one of the essential features in the calculation workflow module because of complex and diversified computing task requirements. 
We drew on the factory method in the software design pattern. We designed many calculation templates of crucial properties, especially the properties related to optoelectronic materials. 
In HT calculations, these property calculations can be arbitrarily combined by adding their name to the tasks array in the task control file named ``input.py''. Moreover, we designed a self-defined template to meet the needs of the customized calculation tasks. 
Users only need to write a small amount of code and can implement their customized calculations. To implement one customized calculation, users need to add the name of this task in the ``input.py'' file and prepare a workflow file with this task's name. 
The tasks module can automatically detect this customized calculation located in the ``diyflow'' directory and perform the calculations by calling its diy\_calculator() method. 
The HT calculations involve the configuration of computing resources. 
In JAMIP, all configuration information is stored in the YAML files in the ``env'' dictionary. 
We have built-in configuration templates for different job management systems and compiler environments. Currently, JAMIP supports three job management systems: PBS, LSF, and SLURM. 
The code provides currently complete interface support for the Vienna Ab initio Simulation Package (VASP)\cite{kresse_efficient_1996}, partial interface support for Quantum Espresso\cite{giannozzi_quantum_2009}, and for more widely used ab initio codes in future releases.

2.1.3 Material property calculations

JAMIP contains a large number of built-in computational templates related to the semiconductor's properties, including thermodynamic, geometrical, electronic, mechanical, phononic, and photonic properties (listed in Table II). 
Some of these calculation tasks interact with the other modules of the program, such as the JAMIP's database and materials production factory. 
Considering the calculation of the decomposition pathway as an example, the program can automatically search for crystallographic information of competing phases related to this material in the JAMIP database, calculate all possible decomposition paths in batches, and output visual analysis results. 
Most importantly, these calculations can be combined arbitrarily into one HT calculation task, in which it can simultaneously calculate multiple properties and handle their calculated results. Furthermore, as discussed above, JAMIP can easily be extended to new customized calculation tasks. 
Note that the above calculations need to be combined with VASP. For Quantum Espresso, the current program only provides some basic calculations, including structural optimization, self-consistent field (SCF) calculation, and electronic band structure calculation. We are continuously expanding the items for property calculations.

\begin{table*}[h]
\centering
\caption{Built-in property calculations.}
\begin{tabular}{l|l}
\hline
Properties & Specific terms\\ \hline
Thermodynamics & Total energy; Enthalpy; Convex hull; Triangle zone; Decomposition of the pathway \\\hline
Geometry & Atomic packing factor; Index of structural connectivity; Tolerance factor; \\ &X-ray diffraction (XRD) pattern; Scanning tunneling microscope (STM) image; \\ &Radial  distribution function (RDF) \\\hline
Electronic structure & Band; DOS/PDOS; Electrostatic potential; Effective mass; Born effective charge;\\ & Deformation potential; Dielectric constant; Electron localization function (ELF); \\ & Charge density; Crystal orbital Hamilton population (COHP/COOP); Bond valence sum\\\hline
Mechanics & Stress tensor; Elastic constants; Bulk modulus; Poisson$^,$s ratio; Hardness \\\hline
Phonon & Phonon dispersion; PHDOS; Raman $\&$ IR; Vibration entropy; Zero-point  energy  \\ &(ZPE);  Thermal coefficient; Gruneisen parameters; Heat capacity; \\ &Thermal conductivity; Soft mode transition\\\hline
Photon & Optical bandgap; Absorption spectrum; Jointed density of states (JDOS); Dipole \\ & matrix elements (DME); Spectroscopic limited maximum efficiency (SLME); \\ & Exciton binding energy\\\hline
Magnetism & Magnetic moment, Magnetism type\\\hline
Ferroelectrics & Electronic polarization magnitude, Polarization pattern\\
\hline
\end{tabular}
\label{table1}
\end{table*}

2.1.4 Calculations guided by artificial intelligence optimization methods

In certain material systems with complex degrees of freedom in composition or structure variation, it is challenging for the HT calculations to exhaust the materials to be explored and exhibit a combinatorial explosion in numbers. 
To achieve optimal functionality of materials with minimal effort and global sampling in the materials space, we use artificial intelligence optimization methods to guide the material property calculations\cite{franceschetti_inverse_1999, zhang_genetic_2013,zhang_genomic_2012}. 
This approach in nature is the same as in crystal structure search studies that attempted to find the lowest-energy material structures through the intelligent exploration of the potential energy landscape\cite{glass_uspexevolutionary_2006, lonie_xtalopt_2011, pickard_ab_2011,wang_calypso_2012, tipton_grand_2013}. 
Moreover, fitness is the material property rather than the free energy in crystal structure search studies. 
The artificial intelligence optimization methods implemented include genetic algorithm and particle swarm optimization algorithm. 
This allows the users to optimize the target material property in the materials configuration space that needs to be prescribed. 
Such studies are meaningful in exploring the extreme property values that the studied materials can reach, or in studying the material systems with masses of metastable states that can be experimentally realized. 
Otherwise, the stability of the materials needs to be properly considered by taking the free energy as an equally important criterion.

2.1.5 Real-time monitoring and calculation-error management

\begin{figure}[h]
\centering
\includegraphics [width=\columnwidth] {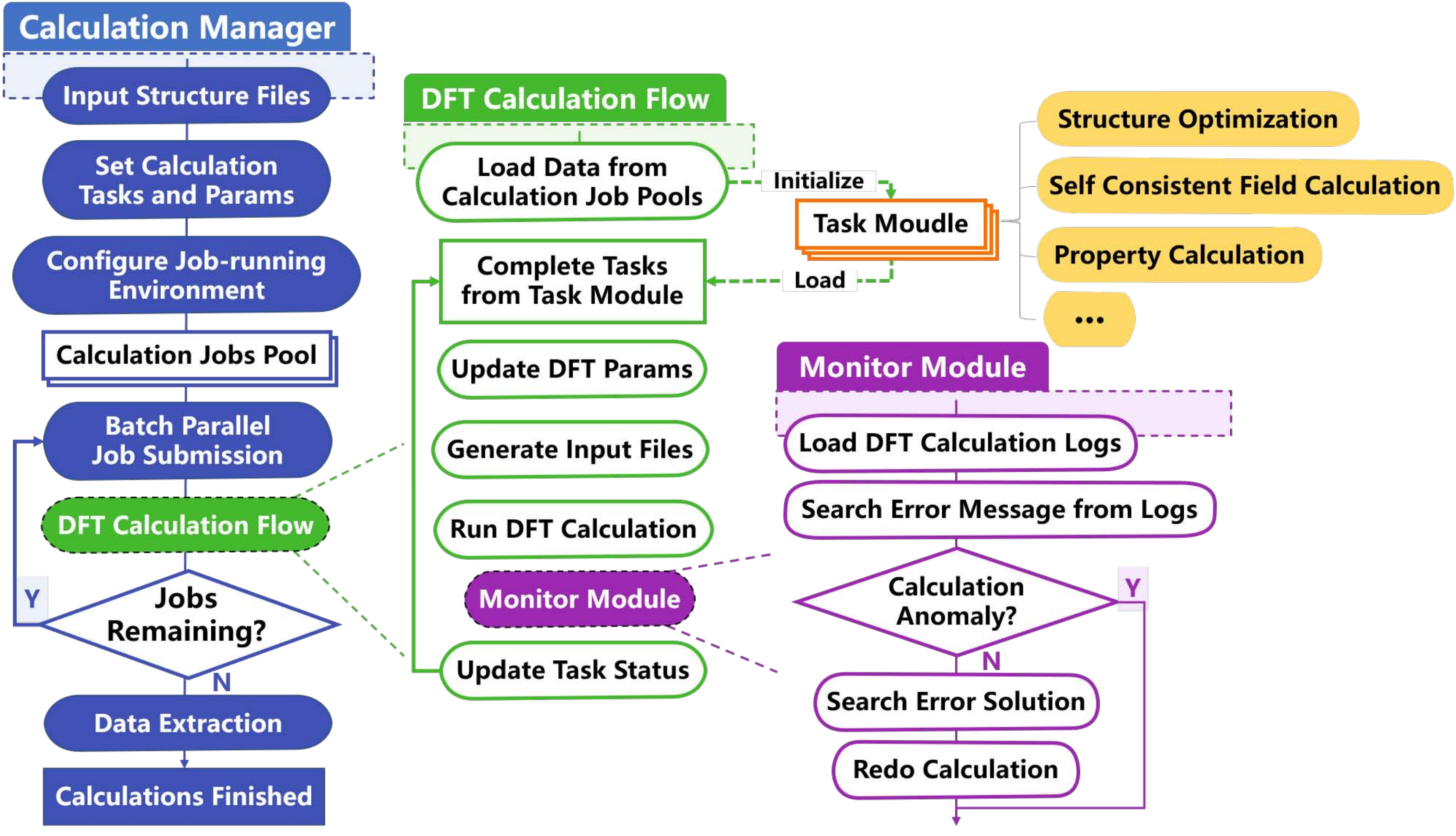}
\caption{The overall perspective of the material database.}
\label{Fig3}
\end{figure}

Adaptive ability relating to the diversified calculation tasks is one of the important indicators for evaluating the quality of HT computational software. 
The default calculation parameters may not be suitable for every sub-calculation task in large-scale calculations owing to the differences in the structure and composition. 
Testing and modifying is a very cumbersome job in the parameter debugging of large-scale calculations. 
It requires a lot of manual intervention and even leads to the failure of tasks in some serious cases. 
In JAMIP, we append the monitoring and self-correction mechanism to deal with run-time errors in the calculations. 
The back-end monitor process is triggered after finishing a calculation task. 
This process judges whether or not the calculation is completed normally. 
For abnormally stopped calculations, the program's control flow automatically switches to the correction step of the error. 
In the current version, we collected some common solutions and constructed an ``error-solution'' database. 
The sketched flow chart is shown in the monitor module of Fig.\ \ref{Fig3}. 
By retrieving this database, JAMIP can automatically try to correct errors encountered in large-scale calculations. 
More importantly, the ``error-solution'' database can be expanded by adding customized solutions of the new error to the YAML-typed file named ``error.yaml'', which comprises three parts: name, keywords, and solution for the error, as shown in Table I. 
With the enrichment of error solutions, the ability of error correction will become more powerful and intelligent and significantly reduce the intensity of human intervention.

\begin{table*}[h]
\caption{Summary of control files.}
\begin{tabular}{l l  l c}
\hline

\multicolumn{2}{l}{\textbf{Calculation params configuration}}      &  \multicolumn{2}{l}{\textbf{Running environment configuration}}\\ \hline
\textbf{base}: & &   \textbf{manager}:PBS&\\
\quad\textbf{ismear}:0 & &  \textbf{queue}:defaut &\\
\quad\textbf{npar}:4 & &  \textbf{cores}:16 &\\
\quad\textbf{sigma}:0.05 & & \textbf{nodes}:1 & \\
\quad\textbf{algo}:fast & & \textbf{maximum}:10 & \\
\textbf{band}: & &\textbf{env}: module load python3&\\
\quad\textbf{lcharg}:11  & & \textbf{...}\\
\textbf{...}: &   \qquad  &&\\\hline
\multicolumn{4}{c}{\textbf{Errors and Corrections configuration}} \\ \hline
\textbf{ISYM}:\\
\quad  - \quad VERY BAD NEWS! internal error in subroutine IBZKPT\\
\quad  - \quad isym: 0\\
\textbf{...}\\\hline
\multicolumn{4}{c}{\textbf{Job submission script}} \\ \hline
1 \quad \textbf{Input}:  Structures, calculation params, ...\\
2 \quad \textbf{Output}: Task Pool\\
3 \quad \textbf{Function}  jump\_input(): \\
4 \qquad  vasp = SetVasp()\\
5 \qquad  vasp.program =`/executable/vasp/program'\\
6 \qquad  vasp.potential=`/directory/of/vasp/pseudopotentials/folders'\\
7 \qquad vasp.task = `scf band partchg emass'\\
8 \qquad vasp.force = 1e-3   &\#force convergence \\
9 \qquad vasp.energy = 1e-6  &\#energy convergence \\
10 \qquad vasp.cutoff = 1.3  &\#cutoff energy of ware function\\
11 \qquad vasp.kpoints = 0.25 &\#k-mesh density\\
\\
12 \qquad pool=Prepare.pool(vasp) \\
13 \qquad pool.set\_structure(`input',`output') &\#read structures\\
14 \qquad pool.save(`silicon.dat',overwrite=True)   &\#save pool of tasks \\
15 \qquad Prepare.cluster(`pbs') &\#check environment configuration\\
16 \qquad Prepare.incar(`vasp.tasks') &\#check calculation configuration\\

\hline
\end{tabular}
\label{table2}
\end{table*}

2.2 Data collection	

JAMIP materials database is based on the Django framework that was developed for database-driven Python web applications. 
Django is an excellent middleware that link the database and application-layered code. 
It can avoid users directly accessing the database by using obscure standard query language (SQL) and reduce the difficulty of development. 
All development activities are in a pure Python environment. 
JAMIP can easily switch to other relational databases. 
Currently, JAMIP supports MySQL, MariaDB, PostgreSQL, Oracle, and SQLite. 
As shown in Fig.\ \ref{Fig3}, the database engine can read crystallographic information from different types of raw structure data files. 
Our developed material production factory can be used as another important source of structures. 
In addition, the database engine can batch-extract calculation results in the data generation stage by executing the command ``jp --db'' in the terminal. 
All the material data are decomposed into some basic data units by the database engine and stored in the underlying relational database.

2.2.1 Data extraction

Once the HT calculations are completed, users need to extract information from different calculation files, summarize/analyze the calculation data, and intuitively obtain the performance differences between materials. 
For instance, the results analysis of band-alignment calculations involves the following: 
(1) finishing HT calculations of the heterojunction, including structural optimization, self-consistent field and band structure calculation; 
(2) extracting the vacuum energy level from the electrostatic potential calculated by self-consistent field and the electronic structure from the output files of band structure calculations, and searching the VBM and CBM of materials by the electron occupation; (3) plotting the band offsets diagram with the aligned band sides, and arranging it in VBM order from low to high. 
We packaged the above data processing flow as a separate post-processor. 
Using this module, users can complete the process of HT calculations, data extraction, and intuitively obtain the relative band edge position and heterojunction type of materials from the band offset diagram.

JAMIP has a complete built-in data extraction module. 
Users can extract the required information from HT output files with the post-processing program and store it as images, or into a comma-separated values (CSV) file and the database. 
The post-processing program currently supports the following functions: 
(1) Total energy, enthalpy, decomposition of the pathway, convex hull, and triangle zone; 
(2) Band structure, electron density of states, charge density, effective mass, born effective charge, deformation potential, dielectric constant, electron localization function, and bond valence sum; 
(3) Stress tensor, elastic constants, bulk modulus, and Poisson's ratio; 
(4) Phonon dispersion, Raman $\&$ IR, vibration entropy, zero-point energy (ZPE), thermal expansion coefficient, thermal conductivity, and soft mode transition; 
(5) Optical bandgap, absorption spectrum, jointed density of states (JDOS), dipole matrix elements (DME), spectroscopic limited maximum efficiency (SLME), and exciton binding energy; \textit{etc.}

2.2.2 Data formatting and normalization

In JAMIP, the calculated raw materials data are extracted and decomposed into a series of Python objects, including entries, structures, compositions, and elements. 
These objects are designed with deep optimizations based on the Django framework. 
They are mainly composed of attributes, associative relationships with other objects, and their auxiliary methods. 
Each object maps to a single database table in our relational database. 
Among these objects, entries are used to save calculation-related information, such as computational parameters and calculated results (\textit{e.g.}, bandgaps and electron/hole effective masses). 
Moreover, each entry object also records its corresponding associated structure. 
Similarly, other objects contain a large number of attributes and association relationships. 
For example, an element object stores a large amount of element-related information (\textit{e.g.}, element symbol, atomic number, and atomic radius) and associated objects such as structures, compositions, and atoms. 
The establishment of these relationships forms a complex relationship network, which is the key to material informatics. 
The formation of these relationship networks also allows users to easily query and filter out the required data by using some simple Django query syntax.

2.2.3 Data storage in the self-contained database

The SQL query is time-consuming because of data filtering and merging operations that are required for large amounts of data entries. 
Optimizing the efficiency of the SQL query is the key to improving the database's performance. 
For example, material structure manipulation involves a large number of queries and operations on the structural information (\textit{e.g}., extracting atoms within a given region and removing these atoms). 
Frequently performing the SQL query operations significantly slows down the response speed of the program. 
Thus, we designed an additional run-time data structure in memory for material structure manipulation. 
Inside a series of Python objects such as structures, compositions, and elements, we maintain a series of attribute arrays synchronously. 
These attribute arrays store all objects associated with them separately. 
When adding or deleting some atoms, all attribute arrays can be refreshed by calling the update() method in the file ``structure.py''. 
By directly accessing these data in memory, the time spent querying the same data is reduced by two orders of magnitude compared to querying the data via the SQL query language. 
However, it should be noted that the query syntax in the run-time data structure is different from the one in the database. 
The key is that only structure objects are allowed to access the run-time data structure. 
For instance, Table III shows the query syntax and efficiency when extracting all atoms of the specified element (Na) in a structure object (s0).

\begin{table*}[ht]
\caption{Query efficiency comparison between the two query methods.}
\begin{tabular}{c c c}
\hline
Type & \quad Query statement & \quad Time-consuming ($\times$10$^-$$^5$s))\\
Run-time data  & s0.get\_element(`Na').atoms &\quad3.4\\
structure database & \quad  Element.objects.get(symbol=`Na').atom\_set.filter(atom\_strucutre=s0) & \quad 290\\
\hline
\end{tabular}
\label{table3}
\end{table*}

\begin{figure}[h]
\centering
\includegraphics [width=\columnwidth] {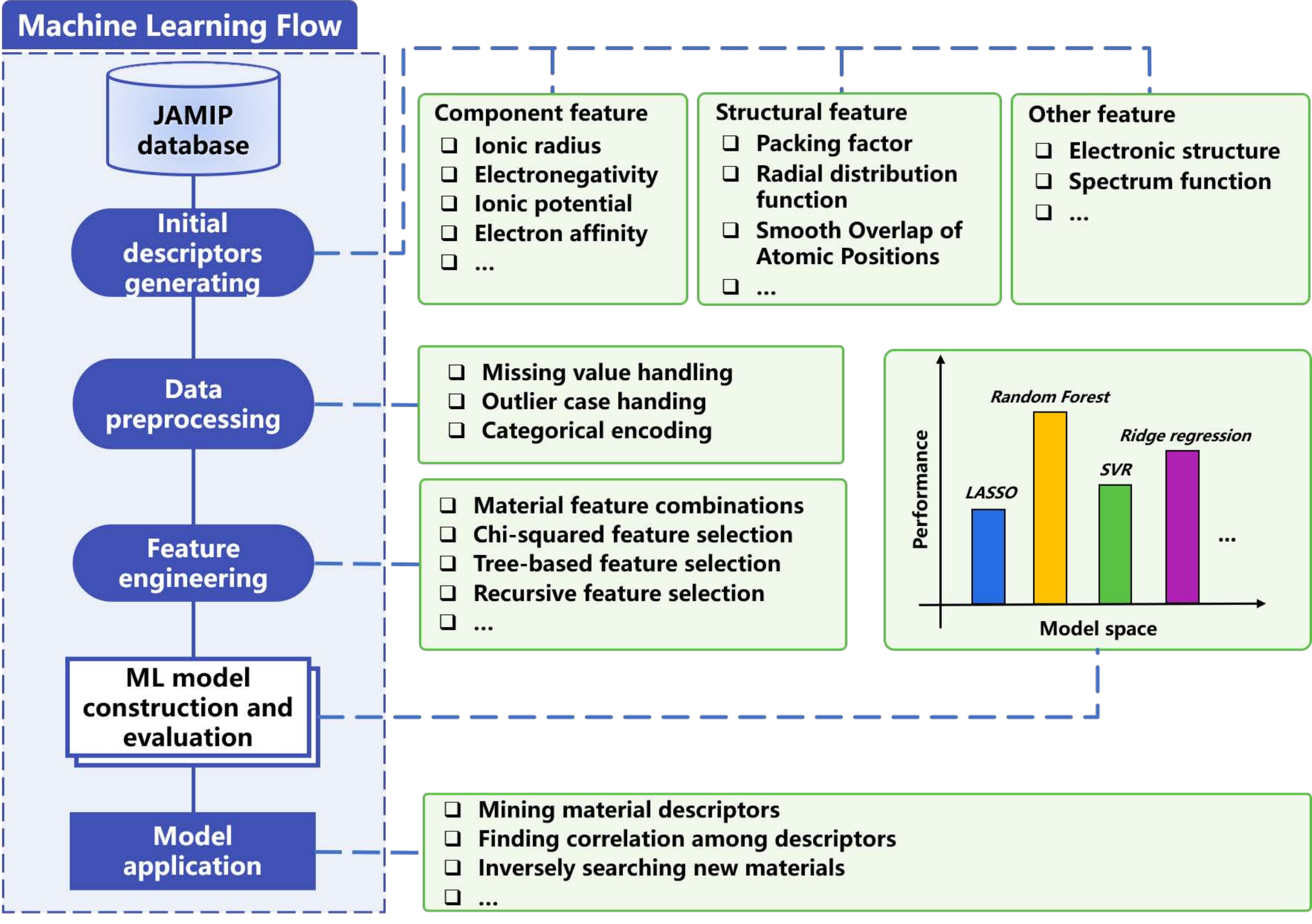}
\caption{Detailed workflow in the Machine-Learning.}
\label{Fig4}
\end{figure}

2.3 Data learning

After completing the HT computing tasks, materials data mining can be performed by studying and learning the data stored in the JAMIP database. 
This version of the JAMIP's data mining module is constructed under the supervised learning model. 
It includes data preprocessing, material feature engineering, machine learning model construction, and performance evaluation (Fig. \ref {Fig4}).

2.3.1 Data filtering and preprocessing

Based on the JAMIP structure classes and database, users can retrieve information on chemical composition (\textit{e.g.}, ionic radius, electronegativity, and ionic potential), crystal structure (\textit{e.g.}, packing factor, radial distribution, and smooth overlap of atomic position), diagrammatic function (\textit{e.g.}, XRD diagram), electronic structure (\textit{e.g.}, band structure and density of state), and optoelectronic properties from structures that have completed the HT tasks. 
Before inputting the data in the machine learning model, the data need to be converted into a language that can be identified by the model for sufficient completion. 
However, there are often missing values or outliers in the building material datasets. 
For example, different structures may have inconsistent dimensions when converted to structural features, or the data distribution may have a ``long tail effect'', JAMIP includes a data cleaning function to deal with missing values and outliers. 
In addition, the original dataset generally has category features, such as crystal systems and space groups. 
It also includes a category feature encoding (\textit{e.g.}, one-hot and binary) function to ensure that the data can be identified.

2.3.2 Material feature engineering

The descriptors in the original dataset are mostly constructed from intuition. 
However, most descriptor sets appear highly dimensional or lack features, which often affects the learning of material properties. Thus, a robust method for identifying descriptors is needed. 
JAMIP's material feature engineering is divided into the following steps: 
1) Feature scaling: This is a normalized data method to ensure that the role of each feature is close to the material target properties (\textit{e.g.}, normalization, non-linear normalization, and regularization); 
2) Feature construction: As the relationship between descriptors and target properties is not simple and is often an exponential or logarithmic relationship, users can customize features by setting the ``expression\_of\_feature'' parameter in the ``feature\_crosses'' method; 
3) Feature selection: After the feature construction is completed and the number of features is added, it also imposes redundant features. 
This increases the model complexity and enhances the overfitting possibility. 
JAMIP offers multiple methods to accomplish feature dimensionality reduction to help uncover the features most closely associated with the material target properties and identify the physical mechanisms behind them, including chi-squared feature selection, tree-based model feature selection, and recursive feature selection.

2.3.3 Machine learning model and performance evaluation

We bulid the machine-learning models by using the scikit-learn machine learning package\cite{franceschetti_inverse_1999}. 
Currently, JAMIP provides four types of machine learning algorithms: regularization linear models, clustering models, kernel models, and tree-based models, and provides post-fitting functions for the above algorithms. 
For different learning tasks (classification or regression), the user can select different evaluation metrics (\textit{e.g.}, root mean square error and confusion matrix) to test the fitted model. 
For randomness in the integrated model, the program takes multiple fittings and averages them to eliminate as much randomness as possible.

\section{Demonstration of usage}

In principle, our JAMIP code can be used to perform materials-informatics-related research on the functional material system of which the functionality-involved properties can be accurately described by first-principles calculations. 
In this section, we briefly describe its use in studying halide perovskites-based semiconductors for optoelectronic applications. 
Halide perovskites (HPs, formula as ABX$_3$), as promising optoelectronic materials (\textit{e.g.}, photovoltaic solar cells, light-emitting diodes, lasers, and photodetectors), have attracted tremendous research interest. 
We chose this family of optoelectronic semiconductors as instances.

\begin{figure}[h]
\centering
\includegraphics [width=\columnwidth] {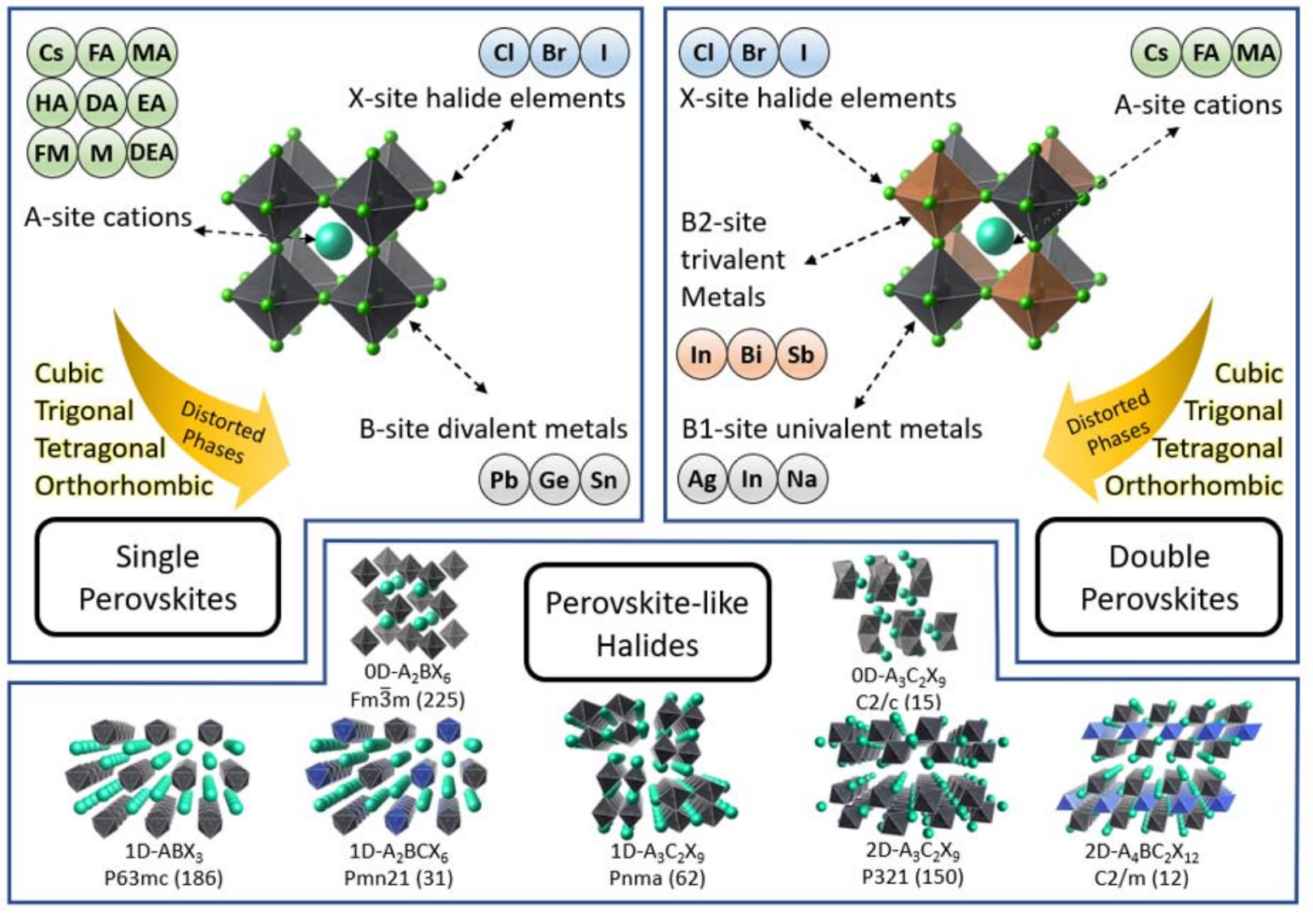}
\caption{Summary of halide perovskites and derivatives in HT calculations, in which ABX$_3$ and A$_2$B$_1$B$_2$X$_6$ (FA=NH$_2$CHNH${_2}{^+}$; MA=CH$_3$NH${_3}{^+}$; HA=NH$_3$OH$^+$; DA=NH$_2$NH${_3}{^+}$; EA=CH$_3$CH$_2$NH${_3}{^+}$; FM=CHONH${_3}{^+}$; M=NH${_4}{^+}$; DEA=CH$_3$NH$_2$CH${_3}{^+}$) for three-dimensional materials, A$_2$BX$_6$ (A=Cs; B=Pd, Sn, Ti; X= Cl, Br, I) and A$_3$C$_2$X$_9$ (A=Cs; C=Bi, In, Sb; X=Cl, Br, I) for 0-dimentional materials, ABX$_3$ (A=Cs, B=Pb, Sn; X=Cl, Br, I), A$_2$BCX$_6$ (A=Cs; B=Ag; C=Bi, In; X=Cl, Br, I), and A$_3$C$_2$X$_8$ (A=Cs; C=Bi, In; X=Cl, Br, I) for one-dimensional materials; A$_3$C$_2$X$_9$ (A=Cs; C=Bi, In; X=Cl, Br, I) and A$_4$BC$_2$X$_{12}$ (A=Cs; B=Cu; C=Bi, In; X=Cl, Br, I) for two-dimensional materials.}
\label{Fig5}
\end{figure}

\begin{figure}[h]
\centering
\includegraphics [width=\columnwidth] {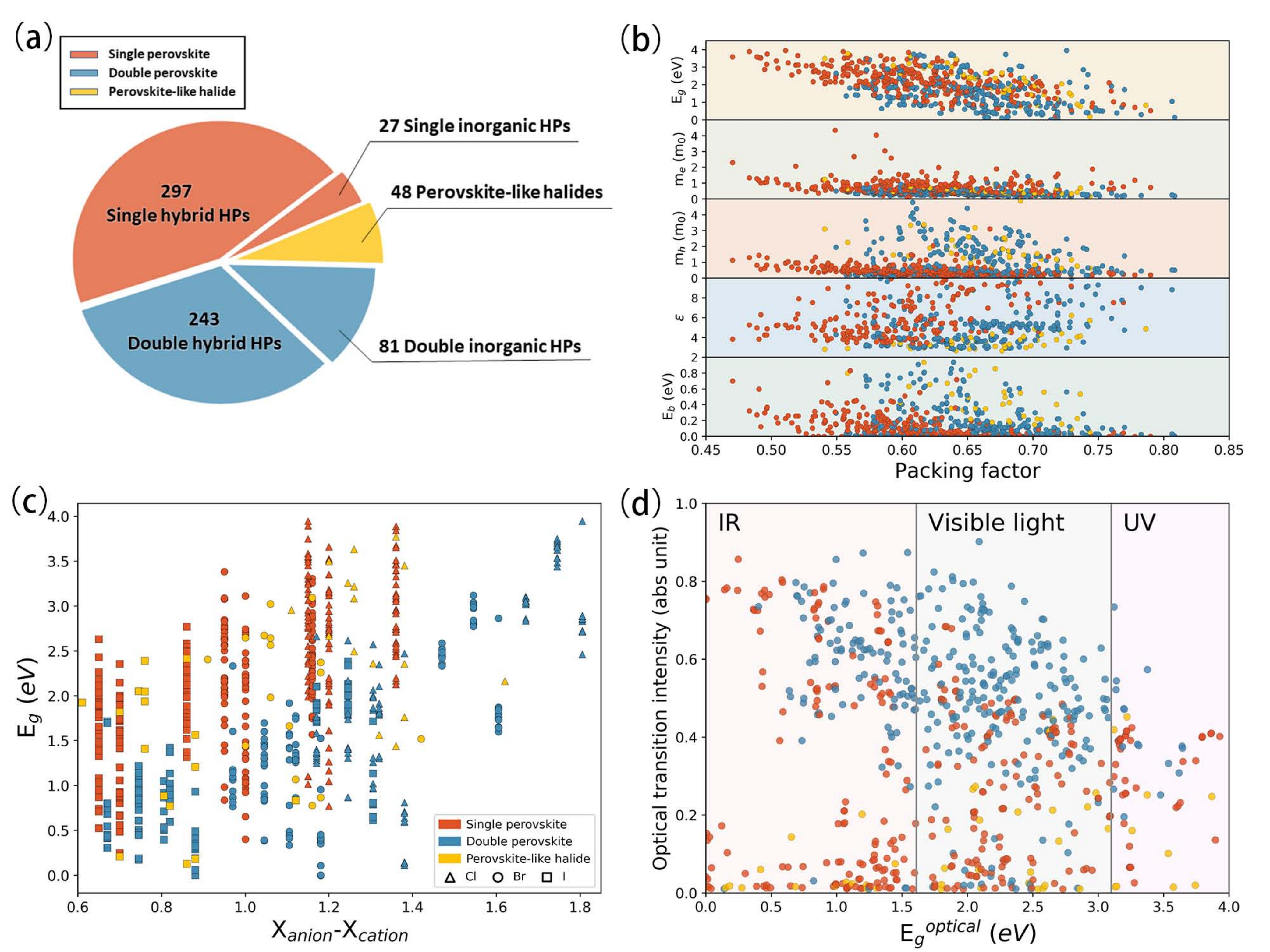}
\caption{Calculated electronic and optical properties of halide perovskites (HPs) and derivatives. 
(a) The number of calculated compounds, including hybrid organic-inorganic perovskites, all-inorganic perovskites, and perovskites-like structures. 
Orange represents the singe perovskites. 
Cyan indicates the double perovskites, and yellow represents the low-dimensional perovskite-like halides. 
(b) Calculated electronic properties (bandgap (E$_g$ in eV), electron and hole effective mass (m$_e$ and m$_h$), dielectric constant ($\varepsilon$), and exciton binding energy (E$_{eb}$ in eV)) versus the atomic packing factor. 
The inorganic atomic radii values were taken from Slater\cite{noauthor_atomic_nodate} and Khan\cite{khan_crystal_1985}, the equivalent atomic radii of organic molecules were taken from Ref.\ \cite{yang2017functionality}. 
(c) Calculated bandgap (E$_g$) versus the electronegative difference between halogen anions and B-site cations. 
The electronegativity values were taken from Pauling's scale\cite{lide2004crc}. 
(d) Inter-band optical transition intensity versus optical bandgap (E${_g}^{optical}$).}
\label{Fig6}
\end{figure}

Fig.\ \ref{Fig5} illustrates the studied materials space of the candidate structures of halide perovskites. 
Besides extensively studied single and double perovskites, we also included some important low-dimensional perovskite-like structures. 
Nearly 700 structures were explored by employing the JAMIP code (Fig.\ \ref{Fig6}a). 
We constructed inputs of three-dimensional candidate structures based on the crystal structure prototypes of single, double, and low-dimensional perovskites with given chemical formula. 
Note that the lattice parameters can be automatically adjusted according to the composition elements when generating the structures. 
The orientation of organic molecules in the candidate structures is generated by using the structure factory method to rotate the molecular orientation appropriately. 
The way to generate the two-dimensional perovskite structure is as follows: (i) we first redefined the crystal lattice of single perovskite structure, rotated the halogen atoms layer by layer to form a twisted octahedron, and (ii) then deleted the outward atoms in the cell and established two-dimensional perovskites.

Next step, by using generated structures as inputs, we carried out the HT calculations of the electronic and optical properties of these halide perovskites with a series of different compositions and structural dimensionality. 
For each candidate structure, we set the automatic calculation flow to work as following: 
(i) firstly perform structural optimization on the perovskite structure and self-consistent field calculation to obtain stable crystal structure and ground-state charge density; 
(ii) calculate the properties of perovskites, including band structure, effective mass, absorption spectrum, dielectric constant, exciton binding energy, and dipole matrix element, etc. 
To facilitate analysis and obtain knowledge, we extracted structure-, energetic-, electronic-, and optical-related information from the HT calculation output files, and imported these information into the JAMIP database. 
Some of the calculated results were shown in Fig.\ \ref {Fig6} as the function of the packing factor (PF) that is an important descriptor to judge the structure stability of perovskites. 
Fig.\ \ref {Fig6}b shows non-monotonic trends of calculated electronic properties in 3D and 2D perovskites (including bandgap, carrier masses, electron-contributed dielectric constant, and exciton binding energy) as the function of the PF, indicating the weak coupling between electronic properties and PF. 
Interestingly, the magnitudes of electron effective masses for different structures distributes in a narrow range, while the magnitudes of hole effective mass distribute a broad range, as well as bandgap, electron-contributed dielectric constant, and exciton binding energy results. 

Fig.\ \ref {Fig6}c illustrates the bandgap values of different halide perovskites as the function of electronegativity difference between anions and cations (\textit{e.g.}, the difference between X and B site in ABX$_3$ halide perovskites). 
For the ionic crystals, the electronegativity difference between anions and cations was regarded as a good descriptor of the bandgap. 
Taking rock salt structures (NaX, X=F, Cl, Br, I) as an example, the bandgap decreases monotonically with the electronegativity difference. 
Indeed the bandgap values of halide perovskites monotonically depend on the electronegativity difference, though the magnitudes of bandgap for single, double, and low-dimensional perovskites distributes in a broad range. 
Fig.\ \ref {Fig6}d shows the optical transition intensity for different candidate structures as function of band gap, where no obvious tendency could be observed. 
The above analyses indicate that understanding the compounds-structure-property relationship is a huge challenge that need researchers' sharp intuition and rich experience.

To facilitate the next machine-learning study and future usage of the data, we stored our calculated data in the database. 
The stored calculation data contain the relaxed structure and calculation instance. 
The calculation instance represents a series of information, including input parameters (such as initial structure and calculation parameters) and extracted output information, such as relaxation structure, calculation time, calculated output energy band data, effective mass, dielectric constant, optical absorption spectrum, and exciton binding energy. 
In the database, there is a one-to-many relationship between the structure model and the calculation instance. 
Users can access the entry properties of structure class to obtain past calculation data.

\begin{figure}[h]
\centering
\includegraphics [width=\columnwidth] {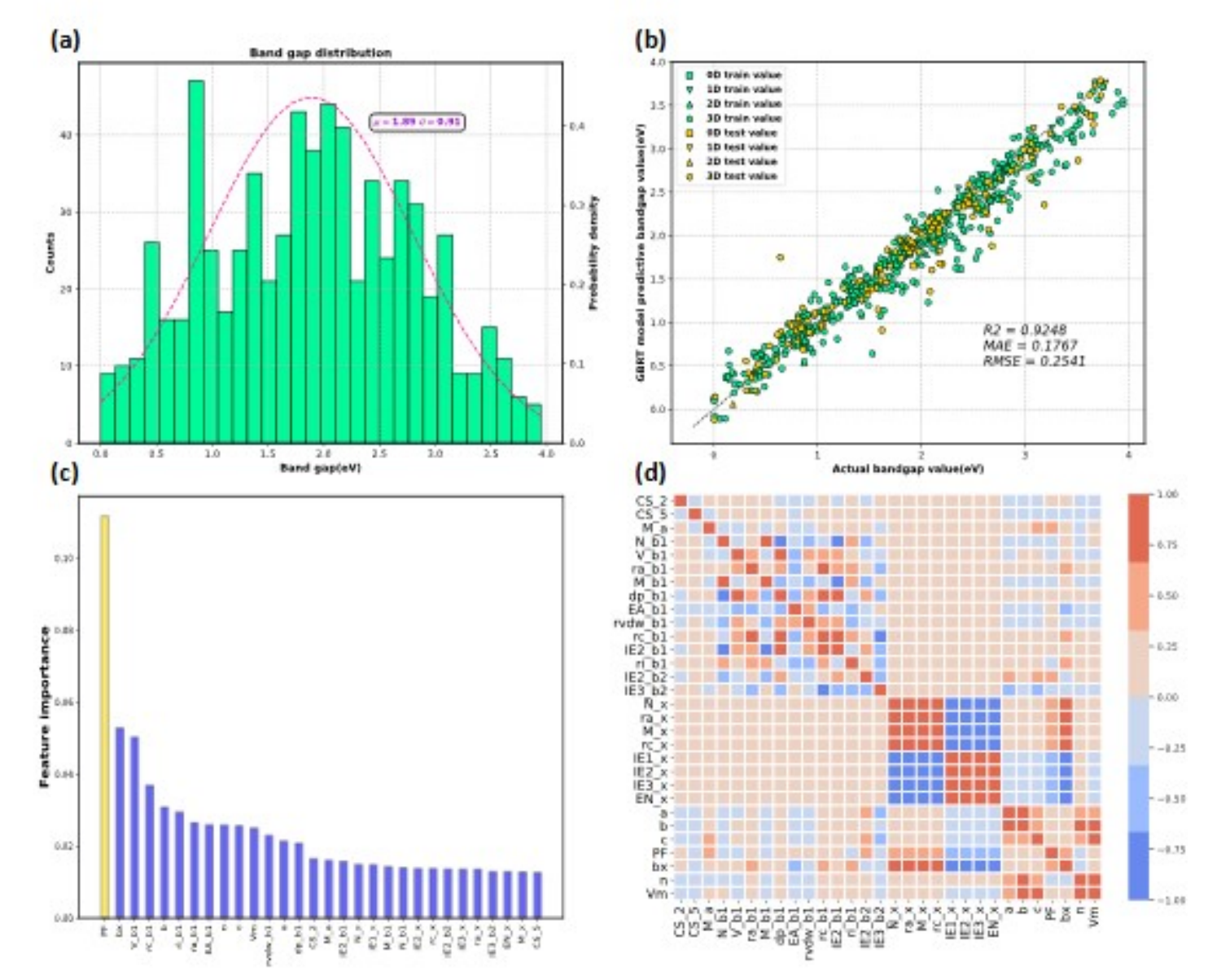}
\caption{Learning the physical rules related to bandgaps. 
(a) Histogram of bandgaps of 696 structures in HT calculations. 
(b) Comparison of the bandgap predicted by the GBRT model with respect to the bandgap by DFT calculations. 
(c) Top 30 important features ranked by GBRT algorithm. 
(d) Heat map of Pearson correlation coefficient matrix among the 30 most important features. 
The blue colors indicate a negative correlation, while the orange colors represent a positive correlation.}
\label{Fig7}
\end{figure}

Machine-learning algorithms combined with big data of materials allow computers to automatically learn the structure-property relationship without human intervention or assistance. 
Here, considering bandgap as an example, we call the gradient boosting regression tree (GBRT) model via the model construction module of JAMIP, which is a tree-based integrated supervised learning algorithm that has many applications in the materials field\cite{lu_accelerated_2018, im_identifying_2019}.
Here, based on calculated halide perovskites, a total of 60 features (Table S2) were generated by calling the JAMIP database to produce the initial dataset. 
Of these features, the crystal system was encoded by calling the JAMIP one-hot encoding method. 
Using statistics of the target values, we found the bandgaps calculated by the density functional theory (DFT) methods obey Gaussian distribution ($\mu$=1.89, $\sigma$=0.91). 
To evaluate the effectiveness of the training model, we divided the dataset into 80\% for training and 20\% for testing. 
In the training process, we used Bayesian optimization combined with the cross-validation approach to finding the optimal hyperparameters. 
The coefficient of determination (R$^2$), mean absolute error (MAE), and root mean square error (RMSE) were then used to evaluate the fitting quality of the hyperparameters by the JAMIP's model evaluation module. 
After several iterations, the GBRT model achieved high fitting accuracy, with 0.9256 of R$^2$, 0.1750 of MAE, and 0.2526 of RMSE (Fig.\ \ref{Fig7}b). 
This indicates by choosing an appropriate machine learning model and enough features, one can achieve a rather accurate description of the bandgap values in the current halide perovskites materials. 
Most importantly, we then used the JAMIP tree-based model feature selection method to mine the features that are most relevant to the target property bandgap, as shown in Fig.\ \ref{Fig7}c. 
We found that the packing factor plays the most important role in determining the values of bandgaps. 
The element information related to B-site is more important than A/X-site. 
This is consistent with the previously reported results. 
We also used the Pearson correlation coefficient matrix to identify the correlations between these features (Fig.\ \ref{Fig7}d). 
We found that most features exhibit low correlation character. 
This implicates that alternative effective features may be needed to reach a good description of the electronic structures of the current complex materials system.

\section{Conclusion and discussion}

To summarize, we have reported the development of an artificial-intelligence-aided data-driven infrastructure named Jilin Artificial-intelligence aided Materials-design Integrated Package (JAMIP), which is designed purposely to meet the requirements of the studies of computational materials informatics. 
It is an open-source software implemented mainly in Python language. 
With the emphasis on automation, extensibility, reliability, and intelligence, it is integrated by intimately connected units of Data generation (\textit{e.g.}, high-throughput materials calculations and screening), Data collection (\textit{e.g.}, automatic data extraction, management, and storage), and Data learning (\textit{e.g.}, integrated descriptors engineering, and machine learning-based data mining functions). 
We integrated the essential modules into the software for computational materials informatics including materials production factory, high-throughput first-principles calculations engine, computational tasks monitoring progress, data extraction, management and storage system, and artificial intelligence data mining and machine learning function. 
The materials production factory consists of specially designed material structure manipulation operators and an inorganic crystal structure prototype database, which facilitates batch processing of materials production for high-throughput calculations. 
In the high-throughput first-principles calculations engine we have packaged various material property calculation functions ranging from thermodynamic, geometrical, electronic, mechanical, phononic, photonic to magnetic properties. 
This is expected to be beneficial to diverse functional materials discovery and design. 
We have designed a computational task allocation and monitoring system to guarantee efficient and reliable large-scale calculations. 
We have included in the software the computational data extraction and analysis toolkits, as well as a deliberately developed database for data storage based on the Django framework. 
We have integrated the artificial intelligence machine-learning based data mining functions that cover data preprocessing, material feature engineering, machine learning model construction, and performance evaluation. 
We demonstrated how our software can be used in performing materials informatics studies of optoelectronic semiconductors by taking halide perovskites as an instance. 

The challenges still faced in the computational materials informatics field are motivating us to further develop our software. 
Firstly, the insufficiency in computational materials data is usually one of the most severe challenges in data mining studies. 
To relieve this challenge, we are developing more material structure manipulation operators into the materials production factory, and meanwhile are working on an efficient structure search algorithm based on machine learning approaches. 
Secondly, the huge computational cost is a serious problem during data generation through high-throughput first-principles calculations, especially for complex systems with hundreds or even thousands of atoms. 
On this aspect, we are launching in the infrastructure our specially developed machine-learning potential for energetic calculations with good transferability. 
We have also been developing efficient and meanwhile accurate electronic structure calculation approach based on machine learning. 
Thirdly, seeking effective descriptors in machine-learning based studies is one of the most important subjects in computational materials informatics. 
We are developing alternative descriptors to accurately represent the structural, electronic structure, and spectral, etc. properties of materials. 
Fourthly, evaluating the stability of designed materials is an essential step when new materials were designed theoretically and call for experimental verification. 
In addition to the thermodynamic criteria implemented in the software, we are developing other approaches associated with kinetic processes. 
This involves finding effective algorithms to calculate the transition barriers of chemical reaction or phase transformation, as well as the algorithms to simulate the crystal growth procedure. 
We believe that with continuous development and improvement, our open-source JAMIP software represents an important addition to the facilities/infrastructures needed by computational materials informatics, and will contribute to accelerating new materials discovery and design from the computational side.

\acknowledgements
This work is supported by the National Natural Science Foundation of China (Grants No. 61722403, 92061113 and 12004131). 
Calculations were performed in part at the high-performance computing center of Jilin University.

\bibliography{Manuscript-3-14}

\end{document}